%
%
\magnification=1200
\baselineskip=15pt
\def\F{{\cal F}}

\def\d{{\rm d}}
\def\l{\,{\rm log\,}}
\def\al{\bar a_l}
\def\ak{\bar a_k}

\def\x1{x_1^-}
\def\12{{1\over 2}}

\rightline{UCLA/96/TEP/32}
\bigskip
\centerline{\bf THE EFFECTIVE PREPOTENTIAL OF N=2 SUPERSYMMETRIC}
\bigskip
\centerline{{\bf SO($N_c$) AND Sp($N_c$) GAUGE THEORIES}
\footnote*{Research supported in part by the National Science 
Foundation under grants PHY-95-31023 and DMS-95-05399}}
\bigskip
\bigskip
\centerline{{\bf Eric D'Hoker}${}^1$, 									{\bf I.M. Krichever}${}^2$,
            {\bf and D.H. Phong} ${}^3$}
\bigskip
\centerline{${}^1$ Department of Physics}
\centerline{University of California, Los Angeles, CA 90024, USA}
\centerline{e-mail: dhoker@physics.ucla.edu}
\bigskip 
\centerline{${}^2$ Landau Institute for Theoretical Physics}
\centerline{Moscow 117940, Russia}
\centerline{and} 
\centerline{Department of Mathematics}
\centerline{Columbia University, New York, NY 10027, USA}
\centerline{e-mail: krichev@math.columbia.edu}
\bigskip
\centerline{${}^3$ Department of Mathematics}
\centerline{Columbia University, New York, NY 10027, USA}
\centerline{e-mail: phong@math.columbia.edu}

\bigskip
\bigskip
\centerline{\bf ABSTRACT}
\bigskip

We calculate the effective prepotentials for N=2 supersymmetric
SO($N_c$) and Sp($N_c$) gauge theories, with an arbitrary number of 
hypermultiplets in the defining representation, from restrictions of the
prepotentials for suitable N=2 supersymmetric gauge theories with
unitary gauge groups. (This extends previous work in which the
prepotential for N=2 supersymmetric SU($N_c$) gauge theories was
evaluated from the exact solution constructed out of spectral curves.)
The prepotentials have to all orders the logarithmic singularities of
the one-loop perturbative corrections, as expected from
non-renormalization theorems. We evaluate explicitly the contributions
of one- and two-instanton processes.

\vfill\break 

\centerline{\bf I. INTRODUCTION}

\bigskip

Powerful techniques are now available for the evaluation of the
effective prepotential of N=2 supersymmetric Yang Mills theories in
their Abelian Coulomb phase (where the gauge group is broken down to
an Abelian subgroup). The effective prepotential, as well as the
masses of the BPS states, are determined from a spectral curve, together
with a meromorphic 1-form $d\lambda$, both of which are parametrized by
the vacuum expectation values of the scalar fields (also called order
parameters). The original developments for an SU(2) gauge group are
in [1], the spectral curve and meromorphic 1-form were determined for
other gauge groups in [2, 3, 4, 5], and the effect of $N_f$
hypermultiplets in the fundamental representation were also included
for SU($N_c$) gauge groups in [6], [7].
\medskip
In a recent paper [8], we developed methods for determining the
prepotential from the spectral curves for arbitrary SU($N_c$) gauge
group  and arbitrary numbers of hypermultiplets $N_f < 2N_c$, in the
regime where the renormalization scale $\Lambda$ is small. We explicitly
calculated the full expansion of the renormalized order parameters
(obtained from the A-periods of $d\lambda$) using the method of
residues, and provided a simple and systematic algorithm for the
evaluation of the renormalized dual order parameters (obtained from the
B-periods of $d\lambda$). Using these methods, we confirmed N=2
supersymmetry non-renormalization theorems and worked out explicitly
the perturbative corrections as well as the 1- and 2-instanton
contributions to the effective potential. These results were found to
agree with those of [1] for SU(2), with those of [9] for
SU(3), as well as with direct field theory calculations in [10]
for SU(2) with $N_f<4$ hypermultiplets in the fundamental
representation  and in [11] for SU($N_c$) with $N_f=0$, both to 1
instanton order. We also showed that the different models [6, 7, 12] for
the spectral curves that were proposed for the cases $N_f \geq N_c+2$,
all give rise to the same effective prepotential.
\medskip
In the present paper, we extend the above results to the cases of all
classical groups, including SO($N_c$) and Sp($N_c$), with any number
of hypermultiplets so as to keep the theory asymptotically free. We
make use of the fact that the spectral curves associated with the
classical groups SO($N_c$) and Sp($N_c$) are hyperelliptic, and may be
viewed as restrictions of the spectral curves for SU($N_c$).
\footnote{*}{
For the gauge group Sp($N_c$), the identification of its
spectral curve with a restriction of a curve for a unitary group
appears possible only when there are at least two exactly massless
hypermultiplets in the defining representation of Sp($N_c$). As we shall
see, this condition appears in our work for purely technical reasons;
it is unclear to us at this point whether it is in any way 
fundamental.} 
Analogously, we show that the homology cycles, the meromorphic 1-form,
and thus the entire effective prepotential may be obtained by simple
restriction from the unitary case. These results imply that, to all
orders in the instanton expansion, all logarithmic singularities of the
prepotential are just those of one loop perturbation theory, thereby
confirming the N=2 supersymmetry non-renormalization theorems. Also,
they show that the prepotential is unchanged under analytic
redefinitions of the classical order parameters, just as we showed for
the case of SU($N_c$) in [8].
\medskip
For the gauge groups SO($N_c$) and Sp($N_c$), we shall work out
explicitly the perturbative corrections as well as the contributions
of 1- and 2-instanton processes to the prepotential and arbitary
numbers of hypermultiplets in the defining representation of the color
group (see however the previous footnote), with the restriction that
the theory remain asymptotically free.

\bigskip

\centerline{\bf II. SPECTRAL CURVES, 1-FORMS AND HOMOLOGY CYCLES}

\bigskip

We consider N=2 supersymmetric gauge theories with classical gauge
groups SU($r+1$), SO($2r+1$), Sp($2r$) and SO($2r$), which  are all of
rank $r$, and numbers of colors $N_c =r+1,~ 2r+1,~2r$ and $2r$
respectively. We also assume that there are $N_f$ hypermultiplets,
transforming under the defining representation of the gauge group, of
dimension $N_c$, and with bare masses $m_j, ~ j=1, \cdots, N_f$.
The N=2 chiral multiplet contains a complex scalar field $\phi$ in the
adjoint representation of the gauge group. The flat directions in the
potential correspond to $[\phi,\phi^{\dagger}]=0$, so that the classical
moduli space of vacua is $r$-dimensional, and can be parametrized
by the eigenvalues $\ak,~ k=1,\cdots, r$ of $\phi$, in the following
way
$$
\eqalign{
{\rm SU}(r+1) 
\qquad \quad
\phi &= {\rm diagonal} ~[\bar a_1, \cdots , \bar a_r, \bar a_{r+1}]
\qquad \qquad 
\bar a_1 + \cdots + \bar a_r + \bar a_{r+1} =0
\cr
{\rm SO}(2r+1) 
\qquad \quad
\phi &= {\rm diagonal} ~[{\cal A}_1, \cdots, {\cal A}_r,0]
\cr
{\rm Sp}(2r) 
\qquad \quad
\phi &={\rm diagonal} ~[\bar a_1,-\bar a_1,\cdots ,\bar a_r ,-\bar a_r]
\cr
SO(2r) 
\qquad \quad
\phi &= {\rm diagonal} ~[{\cal A}_1, \cdots, {\cal A}_r]
\qquad\qquad\qquad 
{\cal A}_k = \pmatrix{ 0 & \ak \cr -\ak & 0 \cr}
\cr
}
\eqno(2.1)
$$
For generic $\ak$, the gauge symmetry is broken down to ${\rm U(1)}^r$
and the dynamics of the theory is that of an Abelian Coulomb phase.
The Wilson effective Lagrangian of the quantum theory to leading order
in the low momentum expansion in the Abelian Coulomb phase is of the
form (in N=1 superfield notation)
$$
{\cal L}={\rm Im}{1\over 4\pi}[\int d^4\theta 
{\partial\F(A) \over \partial A^k} A^{\dagger ~k} +
\12\int d^2\theta
{\partial ^2 \F(A) \over\partial A^k\partial A^l} W^kW^l]
\eqno(2.2)
$$
where the $A^k$'s are N=1 chiral superfields whose scalar components 
correspond to the $\bar a_k$'s at the classical level, and $\F$ is the
holomorphic prepotential. 

The Seiberg-Witten Ansatz for the effective prepotential $\F$ is based on
the choice of a fibration of spectral curves over the space
of vacua, and a meromorphic 1-form $\d\lambda$ over each of these curves.
The renormalized order parameters $a_k$'s of the theory, their duals
$a_{D,k}$'s, and the prepotential $\F$ are then given by
$$
2\pi i \,a_k= \oint_{A_k}\d\lambda,\qquad 
2 \pi i \,a_{D,k}= \oint_{B_k}\d\lambda, \qquad
a_{D,k}={\partial\F\over\partial a_k}
\eqno(2.3)
$$   
with $A_k$, $B_k$ a suitable set of homology cycles on the spectral
curves.

For SU($N_c$) gauge theories, with $N_f<2N_c$ hypermultiplets in the
defining representation of the gauge group, general arguments based on
holomorphicity of $\F$, perturbative non-renormalization beyond 1-loop
order, the nature of instanton corrections and the restrictions of
$U(1)_R$ invariance, suggest that $\F$ should have the following form
\footnote{*}{We shall omit contributions to $\F$ that are of the form
of a $\Lambda$-independent constant times the classical prepotential $\sum
_k a_k^2$ throughout this paper.} 
$$
\eqalignno{  
\F  _{SU(N_c) ;N_f} & (a_1,\cdots  a_{N_c}; m_1,\cdots, m_{N_f}; \Lambda)
\cr
 = &
-{1\over 8\pi
i}\big(\sum_{k,l=1}^{N_c}(a_k-a_l)^2\l{(a_k-a_l)^2\over\Lambda^2}
-\sum_{k=1}^{N_c}\sum_{j=1}^{N_f}(a_k+m_j)^2\l
{(a_k+m_j)^2\over\Lambda^2}\big)\cr 
&+
\sum_{d=1}^{\infty}
\F ^{(d)} _{SU(N_c);N_f} (a_1,  \cdots , a_{N_c}; m_1, \cdots ,
m_{N_f};\Lambda) 
&(2.4)\cr}
$$
The terms on the right hand side are respectively the contribution of
perturbative one-loop effects (higher loops do not contribute in view of
perturbative non-renormalization theorems), and the contributions of 
$d$-instantons
processes. The results for $d=1$ and $d=2$ were computed explicitly in
[8], and we shall record them here for later reference
$$
\eqalign{
\F^{(1)}=&{1\over 8\pi i}\Lambda^{2N_c-N_f}\sum_{k=1}^{N_c}S_k(a_k)
\cr
\F^{(2)}=&{1\over 32 \pi i}\Lambda^{2(2N_c-N_f)}\biggl [ ~
\sum_{k\not=l}{S_k(a_k)S_l(a_l)\over (a_k-a_l)^2}
+{1\over 4}\sum_{k=1}^{N_c}S_k(a_k){\partial ^ 2S_k (x)\over\partial
x^2} \big | _{x=a_k} ~\biggr ]
\cr}
\eqno(2.5)
$$
where the fundamental function $S_k(x)$ is defined by
$$
S_k(x) = {\prod _{j=1} ^{N_f} (x+m_j) \over \prod _{l\not=k}
(x-a_l)^2}.
\eqno (2.6)
$$
By construction, these contributions to $\F$ are
invariant under the group of permutations of the variables $a_k$, i.e.
under the Weyl group of $SU(N_c)$. It is of course possible, though in
general cumbersome, to re-express these results in terms of symmetric
polynomials in the variables $a_k$.

\bigskip

\noindent{\bf a) Spectral curves and associated meromorphic 1-form}

\medskip

The spectral curves for the classical gauge groups were derived in [1]
for SU(2), in [2, 6, 7, 12] for general SU($N_c$), in [4, 12] 
for SO($2r+1$)
in [5, 12] for SO($2r$), and in [12] for Sp($2r$). All these curves are
hyperelliptic. In some cases, different curves have been proposed for
the same gauge group and the same hypermultiplet contents. For example,
in the case of SU($N_c$) gauge group and $N_f > N_c+1$ hypermultiplets,
the curves proposed in [6], in [7] and in [12] are all different. 
However, we have shown in [8], by general arguments and confirmed by
explicit calculations up to 2 instanton processes, that the
corresponding effective prepotentials are the same for each of these
different models of curves. This equivalence results from the fact that
the effective prepotential is unchanged under analytic
reparametrizations of the classical order parameters. Also, we note
that for non-simply laced groups, like Sp($2r$), non-hyperelliptic
curves were proposed in [3]. 

For all N=2 supersymmetric gauge theories based on classical groups,
and with $N_f$  hypermultiplets in the defining representation of the
gauge group, hyperelliptic spectral curves with associated meromorphic
1-forms have been proposed as follows
$$
\eqalign{
y^2
&=A^2(x)-B(x)\cr
\d\lambda&={x ~\d x\over y}\bigg(A'-\12 A{B'\over B}\bigg)
\cr}
\eqno(2.7)
$$
Here, $A(x)$ and $B(x)$ are polynomials in $x$, whose coefficients
vary with the physical parameters of the theory, and are given by
$$
\eqalign{
{\rm SU}(r+1) 
\qquad \quad
A(x) 
 &=
\prod _{k=1} ^{r+1} (x-\ak ),
\qquad \qquad \qquad  
\ B(x) = \Lambda ^{2r+2-N_f} \prod _{j=1} ^{N_f} (x + m_j)
\cr
{\rm SO}(2r+1) 
\qquad \quad
A(x)
& = \prod _{k=1} ^r (x^2 - \ak ^2),
\qquad \qquad \qquad
B(x) = \Lambda ^{4r-2N_f -2} x^2 \prod _{j=1} ^{N_f} (x^2 - m_j ^2)
\cr
{\rm Sp}(2r) 
\qquad \quad
A(x)
&=
x^2 \prod _{k=1} ^r (x^2 - \ak ^2) + A_0, 
\qquad
\ B(x) = \Lambda ^{4r -2N_f +4} \prod _{j=1}^{N_f} (x^2 - m_j^2)
\cr
SO(2r) 
\qquad \quad
A(x)
&=
\prod _{k=1} ^r (x^2 - \ak ^2),
\qquad \qquad \qquad
B(x) = \Lambda ^{4r-2N_f -4} x^4 \prod _{j=1} ^{N_f} (x^2 - m_j ^2) 
\cr
}
\eqno (2.8)
$$
where $A_0 =\Lambda ^{2r -N_f +2} \prod _{j=1} ^{N_f} m_j$. 

Notice that the differential $\d\lambda$ only depends upon the ratio
$B(x)/A(x)^2$, so that simultaneous {\it rescaling} of $A(x)$ by a
function $f(x)$ and $B(x)$ by the function $f(x)^2$ leaves the variables
$a_k$ and $a_{D,k}$, and hence the effective prepotential $\F$
invariant. 

\bigskip

\noindent {\bf b) The case of Sp($N_c$) gauge theories}

\medskip

It is apparent from the form of the functions $A(x)$ above that the
case of Sp($N_c$) gauge group is special : there appears an extra
constant $A_0$ that was not present for the other classical groups.
The methods that we shall present do not seem to extend easily to the
case when $A_0 \not=0$, because there is no natural map onto the curve
for unitary groups. Thus, in this paper, we shall restrict analysis to
the case where at least one of the hypermultiplets of the Sp($N_c$)
supersymmetric gauge theory has exactly zero mass. We shall denote
this restricted case by Sp$(N_c)'$. Under this assumption, $A_0=0$ and
using the rescaling property of the prepotential explained in the
previous paragraph, we find that the curve for Sp$(N_c)'$, i.e.
Sp($N_c$) with at least one hypermultiplet of exactly zero mass is
given by
$$
\eqalign{
{\rm Sp}(2r)' 
\qquad \qquad \qquad
A(x) &= x \prod _{k=1} ^r (x^2 - \ak ^2) \cr
(m_{N_f}=0)
\qquad\qquad \qquad
B(x) &= \Lambda ^{4r -2N_f +4} \prod _{j=1}^{N_f -1} (x^2 - m_j^2)\cr
}
\eqno(2.9)
$$
Henceforth, we shall specialize to this case for the gauge group
Sp($N_c$).

Actually, we further notice that when two hypermultiplets are exactly
massless, the rescaled curves for Sp($N_c$) gauge groups admit an even
simpler form, which we shall record here. We denote this case by
Sp$(N_c)''$.
$$
\eqalign{
{\rm Sp}(2r)'' 
\qquad \qquad \qquad
A(x) &=  \prod _{k=1} ^r (x^2 - \ak ^2) \cr
(m_{N_f-1}=m_{N_f}=0)
\qquad\qquad \qquad
B(x) &= \Lambda ^{4r -2N_f +4} \prod _{j=1}^{N_f -2} (x^2 - m_j^2)\cr
}
\eqno(2.10)
$$
These curves have the same genera as the ones for the SO($N_c$) gauge
groups, and their treatment will be carried out completely in parallel
to that of the orthogonal groups. 
\bigskip

\noindent {\bf c) Homology cycles}

\medskip

The hyperelliptic curves for SO($2r+1$), Sp$(2r)''$ and SO($2r$) all
have genus $2r-1$. To each classical root $\ak$, $k=1,\cdots,r$, there
correspond two branch points $x_k ^\pm$, which define a quadratic
branch cut and an associated homology cycle $A _k$
surrounding the cut joining the two branch points. (Due to ${\bf
Z}_2$ symmetry of the curves, under which $x\to -x$, there correspond 
to the negative roots $-a_k$, $k=1,\cdots,r$, two negative branch points
$-x_k ^\pm$, which define a quadratic branch cut and an associated
homology cycle $A _k '$). For the $B_k$ cycle, we choose
the cycle going from $-x_k^-$ to $x_k^-$ in the first sheet,
completed by its counterpart in the second sheet.
We note that $\#(A_k\cap A_l)=\#(B_k\cap B_l)=0$,
$\#(A_k\cap B_l)=\delta_{kl}$, although
$B_k$ intersects also $A'_k$.
The cycles $A_k$ and $B_k$ thus defined are the ones
we shall take for the Seiberg-Witten Ansatz (2.3). 

Taking into account the fact that the
differential $d\lambda$ is itself odd under the ${\bf Z}_2$ symmetry,
under which $x\to -x$, the normalized periods of the differential
$d\lambda$ obtained in this way are 
$$
a_k = {1\over \pi i} \int _{x_k ^-} ^{x_k ^+} \d\lambda,
\qquad \quad
a_{D,k} = {1 \over \pi i} \int _{-x_k^-} ^{x_k^-} \d\lambda,
\qquad \quad
k=1,\cdots,r.
\eqno(2.11)
$$
This normalization is clearly in agreement with the classical limit,
where $\Lambda \to 0$, and $a_k \to \ak$.

\bigskip
\bigskip

\centerline{\bf III. RESTRICTING PREPOTENTIALS FOR UNITARY GAUGE GROUPS}

\bigskip

From the form of the curves for the different gauge groups in (2.7),
(2.8) and restrictions with massless hypermultiplets for the symplectic
groups in (2.9) and (2.10), we see that the curves for the orthogonal
and symplectic gauge groups can be viewed as natural restrictions of
the curves for unitary groups. The precise correspondences are as
follows. 

The curves for SO($2r+1$), Sp$(2r)''$ and SO($2r$) can be obtained from
those of SU($2r$), where the $2r$ classical order parameters of SU($2r$)
are chosen to be $\bar a_1, \cdots ,\bar a_r,-\bar a_1 ,\cdots,-\bar a_r$.
As a result of ${\bf Z}_2$ symmetry, the quantum order parameters $a_k$
then also come in pairs of opposites : $a_1, \cdots , a_r,-a_1, \cdots,
-a_r$. The correspondences of the number of hypermultiplets, $N_f$, in
these theories and their masses is slightly more involved. For
orthogonal groups, the presence of a power of $x^2$ for SO($2r+1$), and
a factor of $x^4$ for  SO($2r$) in the function $B(x)$ in (2.8), forces
us to make identifications with unitary groups with $2N_f+2$ and $2N_f+4$
hypermultiplets of SU($2r$) respectively. For symplectic groups with at
least two massless hypermultiplets, i.e. the case Sp$(2r)''$, the
correspondence is with a theory of $2N_f -4$ hypermultiplets in
SU($2r$). 

The curves for Sp($2r$) without massless hypermultiplets (this includes
the case with no hypermultiplets at all) can be obtained from those of
SU($2r+2$), where the classical order parameters of SU($2r+2$) are
chosen to be $0,0,\bar a_1, \cdots ,\bar a_r,-\bar a_1 ,\cdots, -\bar
a_r$, and the number of SU($2r+2)$ hypermultiplets is $2N_f$. The
appearance of the double zero at $\bar a=0$ implies that the
corresponding SU($2r+2$) theory has an unbroken SU(2) invariance and is
not in the Abelian Coulomb phase at the classical level. The expansion
methods developed in [8] for the effective prepotential do not apply to
this case, and we shall not consider it again in this paper.

\bigskip

\noindent {\bf a) Restriction of the quantum order parameters $a_k$ and
$a_{D,k}$}

\medskip

Given the above restrictions of the curves of unitary gauge groups to
SO($N_c$) and Sp($N_c$), and the fact that the functional form of the
meromorphic 1-form is the same for the various groups, we obtain the
following relations between the quantum order parameters $a_k$ and
$a_{D,k}$. For maximum clarity, we make all dependences completely
explicit, and we let the range of $k$ and $l$ be $1\leq k,l\leq r$. For
SO($2r+1$), we have
$$
\eqalignno{
a_k \big |   _{{\rm SO}(2r+1);N_f} &(\al; m_1, \cdots ,m_{N_f};\Lambda)
\cr
 = &
\ a_k \big | _{{\rm SU}(2r)} (\al ,- \al; m_1,\cdots ,m_{N_f}, -m_1,
\cdots, -m_{N_f}, 0,0;\Lambda)
 \cr
a_{D,k} \big |   _{{\rm SO}(2r+1);N_f}  &(\al; m_1,\cdots ,m_{N_f};\Lambda)
 &(3.1)\cr
 = &
\ a_{D,k} \big | _{{\rm SU}(2r)} (\al, - \al; m_1, \cdots ,m_{N_f}, -m_1,
\cdots, -m_{N_f}, 0,0;\Lambda)
\cr 
& - a_{D,k+r} \big | _{{\rm SU}(2r)} (\al, - \al; m_1, \cdots ,m_{N_f},
-m_1,\cdots,-m_{N_f}, 0,0;\Lambda)
\cr }
$$
For Sp$(2r)''$, we have
$$
\eqalign{
a_k \big |   _{{\rm Sp}(2r);N_f} & (\al; m_1, \cdots
,m_{N_f-2},0,0;\Lambda)
\cr
 = &
\ a_k \big | _{{\rm SU}(2r)} (\al, - \al; m_1,\cdots , m_{N_f-2}, -m_1,
\cdots, -m_{N_f-2};\Lambda)
\cr
a_{D,k} \big |   _{{\rm Sp}(2r);N_f} & (\al; m_1,\cdots ,m_{N_f-2},
0,0;\Lambda)
\cr
 = &
\ a_{D,k} \big | _{{\rm SU}(2r)} (\al, - \al; m_1, \cdots
,m_{N_f-2},-m_1,\cdots,-m_{N_f-2};\Lambda)
\cr 
& -
a_{D,k+r} \big | _{{\rm SU}(2r)} (\al, - \al; m_1, \cdots  ,m_{N_f-2} ,
-m_1, \cdots,-m_{N_f-2};\Lambda)
\cr }
\eqno(3.2)
$$
For SO($2r$), we obtain
$$
\eqalign{
a_k \big |  _{{\rm SO}(2r);N_f}  & (\al; m_1, \cdots ,m_{N_f};\Lambda)
\cr
 = &
\ a_k \big | _{{\rm SU}(2r)} (\al, - \al; m_1,
\cdots ,m_{N_f},-m_1,\cdots,-m_{N_f}, 0,0,0,0;\Lambda)
\cr
a_{D,k} \big |   _{{\rm SO}(2r);N_f} & (\al; m_1, \cdots ,m_{N_f};\Lambda)
\cr
 = &
\ a_{D,k} \big | _{{\rm SU}(2r)} (\al, - \al; m_1,
\cdots ,m_{N_f},-m_1,\cdots,-m_{N_f}, 0,0,0,0;\Lambda)
\cr
& - a_{D,k +r } \big | _{{\rm SU}(2r)} (\al, - \al; m_1,
\cdots ,m_{N_f},-m_1,\cdots,-m_{N_f}, 0,0,0,0;\Lambda)
\cr}
\eqno(3.3)
$$

In [8], an exact formula was derived for the relation between the
quantum order parameters $a_k$ as a function of the classical order
parameters $\ak$ for gauge group SU($N_c$). Using the above
identifications, we easily extend these exact results to the
case of SO($N_c$) and Sp($N_c$) gauge groups. The result is given in
the form of infinite power series expansions in the renormalization
scale $\Lambda$ :
$$
a_k 
= \bar a_k 
+ \sum _{m=1} ^\infty 
{{\bar \Lambda }^{2m} \over 2^{2m} (m!)^2} 
\biggl ( {\partial \over \partial x} \biggr ) ^{2m-1}
\overline{ \Sigma} _k (x) ^m
\bigg | _{x=\ak}
\eqno(3.4)
$$
with the following results
$$
\eqalign{
{\rm SU}(r+1) 
\qquad 
\bar \Lambda 
 &=
\Lambda ^{r+1-N_f/2}
\qquad \quad   
\overline{\Sigma} _k(x) = \prod _{j=1} ^{N_f} (x+m_j) 
\prod _{l\not=k} (x-\al)^{-2}
\cr
{\rm SO}(2r+1) 
\qquad 
\bar \Lambda 
&=
\Lambda ^{2r-1-N_f}
\qquad \quad 
\overline{\Sigma} _k (x) = x^2 (x+\ak)^{-2} 
\prod _{j=1} ^{N_f} (x^2-m_j^2) \prod
_{l\not=k} (x^2-\al^2)^{-2}
\cr
{\rm Sp}(2r)'' 
\qquad 
\bar \Lambda
&=
\Lambda ^{2r+2-N_f}
\qquad \quad
\overline{\Sigma} _k (x) = (x+\ak)^{-2} \prod _{j=1} ^{N_f-2 } (x^2-m_j^2) 
\prod _{l\not=k} (x^2-\al^2)^{-2}
\cr
{\rm SO}(2r) 
\qquad 
\bar \Lambda
&=
\Lambda ^{2r-2-N_f}
\qquad \quad 
\overline{\Sigma} _k (x) = x^4 (x+\ak)^{-2} 
\prod _{j=1} ^{N_f} (x^2-m_j^2) \prod
_{l\not=k} (x^2-\al^2)^{-2}
\cr
}
\eqno (3.5)
$$
In the above expressions, the range of $k$ is just over the independent
variables, and is thus restricted to $k=1,\cdots, r$.

\bigskip

\noindent {\bf b) The effective prepotential}

\medskip
 
Since the renormalized order parameters $a_k$ and $a_{D,k}$ for
SO($N_c$) and Sp($N_c$) gauge groups may both be obtained as
restrictions from the unitary case, it is natural to expect that also the
effective prepotential may be viewed as such a restriction. The
restriction rules for the prepotential turn out to be particularly
simple in view of the fact that the differences $a_{D,k} - a_{D,k+r}$ are
naturally produced by a straightforward restriction of $\F_{{\rm SU}(2r)}$
to the ${\bf Z}_2$ symmetric arrangements for the gauge groups SO($N_c$)
and Sp($N_c$). As a result, we readily deduce the correct
prepotentials for the orthogonal and symplectic groups. For SO($2r+1$),
we have 
$$
\eqalign{
\F&  _{{\rm SO}(2r+1);N_f} (a_1, \cdots , a_r; m_1, \cdots ,m_{N_f};\Lambda)
\cr
& =
\F _{{\rm SU}(2r);2N_f+2} (a_1, \cdots , a_r,-a_1, \cdots, -a_r; m_1,
\cdots ,m_{N_f},-m_1,\cdots,-m_{N_f}, 0,0;\Lambda)\cr}
$$
for Sp($2r$), with at least two massless hypermultiplets, i.e. the case
Sp$(2r)''$, we have
$$
\eqalign{
\F&  _{{\rm Sp}(2r);N_f} (a_1, \cdots , a_r; m_1, \cdots
,m_{N_f-2},0,0;\Lambda)
\cr
& =
\F _{{\rm SU}(2r);2N_f-4} (a_1, \cdots , a_r,-a_1,\cdots, -a_r; m_1,
\cdots ,m_{N_f-2},-m_1,\cdots, -m_{N_f-2} ;\Lambda)\cr}
$$
and, finally, for SO($2r$), we have
$$
\eqalign{
\F&  _{{\rm SO}(2r);N_f} (a_1, \cdots , a_r; m_1, \cdots ,m_{N_f};\Lambda)
\cr
& =
\F _{{\rm SU}(2r);2N_f+4} (a_1, \cdots , a_r,-a_1,\cdots, -a_r; m_1,
\cdots ,m_{N_f},-m_1,\cdots, -m_{N_f},0,0, 0,0;\Lambda)\cr}
$$
From the above restriction rules, it follows that for each of the gauge
groups, the prepotential may be decomposed in a sum over the number of
instantons contributing to the process, just as was the case for
unitary gauge groups in (2.3). We shall denote by $\F ^{(d)}$ the
contribution arising from $d$ instanton processes, and, for $d\geq1$,
these functions depend on $\Lambda$ through a factor of $\bar \Lambda
^{2d}$ where $\bar \Lambda $ was defined for each group in (3.5). The
contribution from zero instantons, i.e. classical plus perturbative
corrections, is denoted by $\F ^{(0)}$. Using the results from [8], and
the above restriction rules, we now have the following results for the
effective prepotential. 
\medskip
The perturbative contributions $\F^{(0)}$ are given as follows. For
gauge groups $G=$ SO($2r+1$), Sp($2r$) with at least two massless
hypermultiplets, i.e. the case Sp$(2r)''$, and SO($2r$) we have the
following formula
$$
\eqalignno{  
\F  _{G ;N_f}  (\al; m_1,\cdots, m_{N_f}; \Lambda)
 = &
{i\over 4\pi } \biggl \{ \sum _{k\not=l}^{r} \sum _{\epsilon =\pm 1}
(a_k+ \epsilon a_l)^2 \l {(a_k + \epsilon a_l)^2\over\Lambda^2}
\cr
&
+\xi \sum _{k=1} ^r a_k ^2 \l {a_k ^2\over\Lambda ^2}
& (3.6)\cr
&
-\sum_{k=1}^{r}\sum_{j=1}^{N_f} \sum _{\epsilon =\pm 1} 
(a_k+ \epsilon m_j)^2 \l {(a_k+ \epsilon m_j)^2\over\Lambda^2} \biggr \}
\cr} 
$$
where the constant $\xi$ takes on the values $\xi =2,~4$ and $0$ for
$G=$ SO($2r+1$), Sp$(2r)''$ (Sp($2r$) with at least two massless 
hypermultiplets), and SO($2r$) respectively. We readily
recognize these numbers from the structure of the corresponding Dynkin
diagrams.
\medskip
The 1-instanton contributions are also readily deduced from the results
of [8], combined with the restriction rules above. The results are most
easily cast in terms of the parameters
$\bar \Lambda$ and the functions
$\Sigma_k(x)$ defined for each group gauge group $G$ = SO($2r+1$), 
Sp$(2r)''$ (Sp($2r$) with at least two massless hypermultiplets), 
and SO($2r$) 
as in (3.5), but with
{\it the classical order parameters $\ak$ replaced
by their renormalized counterparts $a_k$}
$$
\eqalign{
{\rm SU}(r+1) 
\qquad 
\bar \Lambda 
 &=
\Lambda ^{r+1-N_f/2}
\qquad \quad   
{\Sigma} _k(x) = \prod _{j=1} ^{N_f} (x+m_j) \prod _{l\not=k} (x-a_l)^{-2}
\cr
{\rm SO}(2r+1) 
\qquad 
\bar \Lambda 
&=
\Lambda ^{2r-1-N_f}
\qquad \quad 
{\Sigma} _k (x) = x^2 (x+a_k)^{-2} \prod _{j=1} ^{N_f} (x^2-m_j^2) \prod
_{l\not=k} (x^2-a_l^2)^{-2}
\cr
{\rm Sp}(2r)'' 
\qquad 
\bar \Lambda
&=
\Lambda ^{2r+2-N_f}
\qquad \quad
{\Sigma} _k (x) = (x+a_k)^{-2} \prod _{j=1} ^{N_f} (x^2-m_j^2) \prod
_{l\not=k} (x^2-a_l^2)^{-2}
\cr
{\rm SO}(2r) 
\qquad 
\bar \Lambda
&=
\Lambda ^{2r-2-N_f}
\qquad \quad 
{\Sigma} _k (x) = x^4 (x+a_k)^{-2} \prod _{j=1} ^{N_f} (x^2-m_j^2) \prod
_{l\not=k} (x^2-a_l^2)^{-2}
\cr
}
\eqno (3.7)
$$
Then we have
$$
\F^{(1)}_{G;N_f} 
={1\over 4\pi i} \bar \Lambda^2 \sum_{k=1}^{r}\Sigma _k (a_k)
\eqno(3.8)
$$
(Note : this formula does not apply to SU($N_c$) as written, and would
require an extra factor of $\12$.) 
\medskip
Similarly, the 2-instanton contributions may also be worked out, and we
have 
$$
\F^{(2)} _{G;N_f} 
={1\over 16 \pi i} \bar \Lambda^4 \biggl [ ~
\sum_{k\not=l} ^r \sum _{\epsilon = \pm 1} 
{\Sigma _k(a_k) \Sigma _l(a_l) \over (a_k+ \epsilon a_l)^2} 
+{1\over 4} \sum_{k=1}^{r} \Sigma _k(a_k) 
{\partial ^ 2 \Sigma _k (x) \over\partial x^2}
\big | _{x=a_k} ~\biggr ]
\eqno(3.9)
$$
Again, for SU($N_c$), the above formulas requires an extra factor of
$\12$, and a restriction to $\epsilon =-1$.

\bigskip
\bigskip

\noindent {\bf IV. SPECIAL CASES AND DISCUSSION}

\bigskip

We compare briefly now our results with various special 
cases discussed in the literature and obtained either 
directly from the quantum
field theory using instanton calculations, or from the
Seiberg-Witten type approach.
\medskip
The literature on the effective prepotential for
SO($N_c$) and Sp($N_c$) gauge groups is not nearly as
extensive as that for SU($N_c$). In [16], Ito
and Sasakura evaluate the prepotential,
up to 1-instanton order, from both instanton calculations and 
the Seiberg-Witten
approach in the case of {\it pure} N=2 supersymmetric
Yang-Mills (no hypermultiplets). Using
instanton calculations, they propose a formula
for the 1-instanton correction $\F^{(1)}$
for any simple Lie group. Using the Seiberg-Witten
approach, they derive explicitly Picard-Fuchs
equations in the case of rank $\leq 3$,
and rely on the scaling equations of [17].
For SO($2r+1$) and SO($2r$) gauge groups,
our results for $\F^{(1)}$ do specialize to
theirs if we set $N_f$ to be 0.
For Sp($2r$), it is of course not
possible at the present time to compare
the two results, since in the case they consider, there are
no hypermultiplets, while in ours,
we require at least two massless ones.
It is however intriguing that there
is no obvious way of interpolating
between the two types of expressions
that have been put forth.

A few days ago, another preprint [18] appeared,
which also deals with the Seiberg-Witten
approach for classical gauge groups,
up to 1-instanton order.

\bigskip
\bigskip

\centerline{\bf REFERENCES}

\bigskip

\item{[1]} N. Seiberg and E. Witten, Nucl. Phys. {\bf B426} (1994) 19, 
hep-th/9407087; Nucl. Phys. {\bf B431} (1994) 484, hep-th/9408099.

\item{[2]} A. Klemm, W. Lerche, S. Yankielowicz, and S. Theisen,
Phys. Lett. {\bf B344} (1995) 169;\hfil\break
P.C. Argyres and A. Faraggi, Phys. Rev. Lett. {\bf 73} (1995) 3931, 
hep-th/9411057;
\hfil\break
M.R. Douglas and S. Shenker, Nucl. Phys. {\bf B447} (1995) 271, 
hep-th/9503163;\hfil\break
P.C. Argyres, R. Plesser, and A. Shapere, Phys. Rev. Lett. 
{\bf 75} (1995) 1699,
hep-th/9505100;\hfil\break
J. Minahan and D. Nemeshansky, hep-th/9507032;\hfil\break
M. Alishahiha, F. Ardalan, and F. Mansouri, hep-th/9512005\hfil\break
M.R. Abolhasani, M. Alishahiha and A.M. Ghezelbash, hep-th/9606043

\item{[3]}
E. Martinec and N. Warner, hep-th/9509161, hep-th/9511052

\item{[4]}
U.H. Danielsson and B. Sundborg, Phys. Lett. {\bf B358} (1995) 273,
USITP-95-12, UUITP-20/95;
hep-th/9504102.

\item{[5]}
A. Brandhuber and K. Landsteiner, Phys. Lett. {\bf B358} (1995) 73, 
hep-th/9507008.

\item{[6]} A. Hanany and Y. Oz, Nucl. Phys. {\bf B452} (1995) 73,
hep-th/9505075\hfil\break A. Hanany, hep-th/9509176.

\item{[7]} I.M. Krichever and D.H. Phong, hep-th/9604199, to
appear in J. of Differential Geometry;

\item{[8]} E. D'Hoker, I.M. Krichever and D.H. Phong, hep-th/9609041

\item{[9]} A. Klemm, W. Lerche, and S. Theisen,
hep-th/9505150;\hfil\break K. Ito and S.K. Yang, hep-th/9603073

\item{[10]} N. Dorey, V. Khoze, and M. Mattis, hep-th/9606199, 
hep-th/9607202;\hfil\break
Y. Ohta, hep-th/9604051, hep-th/9604059;

\item{[11]} K. Ito and N. Sasakura, SLAC-PUB-KEK-TH-470,
hep-th/9602073.
\item{[12]} P.C. Argyres and A.D. Shapere, hep-th/9509175

\item{[13]} D. Finnell and P. Pouliot, Nucl. Phys. {\bf B453} (1995)
225

\item{[14]} A. Yung, hep-th/9605096;\hfil\break 
F. Fucito and G. Travaglini,
hep-th/9605215

\item{[15]} T. Harano and M. Sato, hep-th/9608060

\item{[16]} K. Ito and N. Sasakura, hep-th/9608054

\item{[17]} M. Matone, Phys. Lett. {\bf B357} (1995) 342;\hfil\break
J. Sonnenschein, S. Theisen, and S. Yankielowicz,
Phys. Lett. {\bf B367} (1996) 145;\hfil\break
T. Eguchi and S. K. Yang, Mod. Phys. Lett. {\bf A11} (1996) 131.

\item{[18]} T. Masuda and H. Suzuki, hep-th/9609065; hep-th/9609066.

\end